\documentstyle[preprint,aps]{revtex}

\newcommand{\bra}[1]{\left\langle #1 \right|}
\newcommand{\ket}[1]{\left| #1 \right\rangle}

\begin{document}

\draft

\title{Comment on ``No core calculations'' of the spectra
       of light nuclei}

\author{T.\ Engeland, M.\ Hjorth-Jensen,
A.\ Holt and E.\ Osnes}

\address{Department of Physics,
University of Oslo, N-0316 Oslo, Norway}

\maketitle

\begin{abstract}
We comment upon a recent work of Zheng {\em et al.} concerning
calculations of spectra of
light nuclei with no core, where an effective
interaction is constructed which spans over several shells.
It is demonstrated that the omission of the particle-particle
ladder diagrams in their calculations, explains the large differences
between results obtained with various model spaces. We use this to
infer that low-order
perturbation theory works well in reproducing the binding energy
of the system we consider.
\end{abstract}


\clearpage

\section{Introduction}
Recently, Zheng {\em et al} \cite{prc48} (hereafter ZBJVM)
have presented an approach
meant to circumvent the notorious intruder state problem. Moreover,
this approach was devised in
order to avoid calculations of complicated
Feynman-Goldstone diagrams which arise in perturbation theory.
It is a well-known fact that the presence of so-called
intruder states
may lead to the divergence of the order-by-order
pertubative expansion
for the effective interaction $H_{\mathrm{eff}}$.
The latter is understood to
be evaluated from perturbative many-body techniques and is defined
within a physically selected model space, which is given by a projection
operator $P$. The remaining degrees of freedom are accounted for by
the perturbative expansion. These degrees of freedom are represented by
a projection operator $Q$, so that $P+Q=1$ and $PQ=0$.
The idea behind the work of
ZBJVM is, through the use of an enlarged model space,
to avoid both the intruder state problem
and that of calculating many perturbative contributions.

In this comment we show that the effective interactions derived
by Zheng {\em et al.} may not be consistent with the underlying
theory for the effective interaction. Our points
are discussed in the next three sections. In
Section II we discuss the omission of particle-particle
ladders diagrams in the calculations of ZBJVM. Section III
critically discusses the use of the starting energy as a
variable. In Section IV a brief discussion of the non-hermiticity
of the effective interaction is also included, and
our conclusions are drawn in Section V.

\section{No-core shell-model calculations with the $G$-matrix}

The first step in the calculations of Ref.\ \cite{prc48}
is to evaluate the
nuclear raction matrix $G$ given by
\begin{equation}
       G=V+V\frac{Q}{\omega - H_0}G,
\end{equation}
where $\omega$ is the unperturbed
energy of the interacting nucleons, and $%
H_0$ is the unperturbed hamiltonian. The operator $Q$, commonly referred to
as the Pauli operator, is a projection operator which prevents the
interacting nucleons from scattering into states occupied by other nucleons.
There are many ways to handle the Pauli operator of Eq.\ (1). Two of these
are demonstrated in Fig.\ \ref{fig:fig1}. In (a) we show the Pauli
operator obtained through the double-partitioned scheme of Ref.\ \cite{kkko76}.
There one has to define a core, given by the boundary $n_1$, which represents
the last single-hole state.
$n_2$ is the last single-particle state of
the model space.

In the calculations of ZBJVM, the Pauli operator is defined
as in (b) of Fig.\ \ref{fig:fig1}. The model space is again limited by
the boundary $n_2$,
but we have no holes. This definition is the first step
in the so-called
``no-core'' approach of Ref.\ \cite{prc48}.
In this work we use the Pauli operator in (b) of Fig.\ \ref{fig:fig1}
and define the model space to consist of the $0s$-, $0p$-, $1s0d$- and
$1p0f$-shells.
We could then, in principle, use the corresponding
$G$-matrix to obtain the eigenvalues within this model space.
The authors of Ref.\ \cite{prc48} are also interested in studying
how important various model spaces are. They therefore calculate the
eigenvalues with smaller spaces first (see Table I
of Ref. \cite{prc48}), say only
the $0s$-shell. However, they use the $G$-matrix
defined with a model space which includes also the $0p$-, the $1s0d$-
and the $1p0f$-shells.
In so doing, they have to include the ladder
diagram of Fig.\ \ref{fig:fig2}, and higher-order ladder
diagrams as well, with intermediate states
from the $0p$-, $1s0d$- and $1p0f$-shells, since they use different
model spaces in the calculations of spectra and the evaluation
of the $G$-matrix, see e.g.\ the discussion
in Ref.\ \cite{kkko76}. This is, to our knowledge, not
done in Ref.\ \cite{prc48}. Actually, we will demonstrate that the
omission of these ladder diagrams explains to a large extent why ZBJVM
obtain rather different results when they compare results
from diagonalizations with one, two and three oscillator shells,
respectively.

To demonstrate our point, we choose a fictitious system to
consist of two particles only, and define various model spaces.
The conclusions apply equally well to systems with more
particles.
Here we choose our hamiltonian $H$ to consist of
\begin{equation}
  H=H_0+G,
\end{equation}
with the unperturbed single-particle energies which define
$H_0$ given by the harmonic oscillator
\begin{equation}
    \varepsilon_{nl} =\left( 2n +l+\frac{3}{2}\right)\hbar\Omega
    +\Delta ,
\end{equation}
where $\Omega$ is the oscillator energy.
Here we set
$\hbar\Omega =14$ MeV.
We add a negative
shift $\Delta=-71$ MeV in order to obtain negative
starting energies only (see the discussion in Section III)
and use, as in Ref.\ \cite{prc48},
a fixed starting  energy, chosen to be $-100$ MeV. This corresponds to
twice the energy of a single particle state in the $0s$-shell, a choice
we made in order to avoid poles in the calculation of ladder diagrams.
As will be discussed in the next section, the use of a fixed starting
energy implies that we have a degenerate
model space, which is rather questionable if the model space spans
over several shells. The choice of a fixed starting energy is also done
in order to avoid the problems with
the non-hermiticity of the effective interaction discussed
in Section IV.
The parameters
of the Bonn B potential in Table A.1 of Ref.\ \cite{mac89}
are used to define the nucleon-nucleon potential $V$.
The Pauli operator is defined as in (b) of Fig.\ \ref{fig:fig1},
with $n_2$ given by the last single-particle state in the
$1p0f$-shell.

The resulting eigenvalues for the lowest lying
$JT=10$  state is shown in Table I.
The most important components in the wave functions
of this state arise from single-particle states
in the $0s$- and $0p$-shells.

The results labelled $G$, include only the $G$-matrix, as done in the work
of ZBJVM. As can be seen from Table I, there
is clearly a large difference (of the order
of $50\%$ or more) between results obtained with a model
space defined by the $0s$-shell only and a model
space which includes all shells up to the $1p0f$-shell.
This qualitative
pattern also agrees with Table I of ZBJVM.
In their conclusions, Zheng {\em et al.}
use this to infer that one needs to take into account
large model spaces, since the binding energies do not stabilize
as functions of the various model spaces\footnote{Note that
we omit any discussions on excited spectra, since these
include in general more and more
complicated configurations as one
increases the model space.}.
We show in Table I that this conclusion is misleading.
The results obtained with the $G$-matrix plus the
two-particle ladder (2P) diagram up to third  order in $G$ (higher-order
terms are negligible)
for the $0s$ model space, show that these results are rather
close to those obtained with $G$ for the model space which
includes all four shells. This demonstrates clearly that
the lack of stabilization in the calculation
of the ground states in Ref.\ \cite{prc48}, is simply
due to the omission of the particle-particle ladder
diagrams.

Note that in our calculations with $G+2P$ for more than one oscillator
shell, we use a degenerate model space, as done by ZBJVM. This means that
if define the model space to include the $0s$-, $0p$- and $1s0d$-shells,
all single-particle states have the same energy. This approximation
explains also why the results of the $G+2P$ calculations differ
slightly from model space to model space. In this sense, the result
obtained with the $0s$-shell only, is the most rigorous one.
This results shows also that low-order perturbation theory
works well in reproducing the lowest $JT=10$ state.

The results with $G$ should also have taken into account
a non-degenerate model space, but here we have tried to follow
ZBJVM as closely as possible. The problems with a non-degenerate
model space are addressed in the next section.

\section{Role of the starting energy}

In principle, the effective interaction
should not depend on the choice of starting energy $\omega$, though,
since an approximation to the perturbation expansion is made,
the effective interaction may depend on $\omega$.
In Table II of Ref.\ \cite{prc48}, it is shown that the excited
spectra depend weakly on $\omega$, whereas  the ground state
of $^6$Li depends strongly on $\omega$. This state varies from
$-23.044$ MeV to $-29.366$ MeV with starting energies between $\omega =20$
and $\omega =38$, respectively.
The authors of Ref.\ \cite{prc48} give
no physical arguments for why one should choose a given
starting energy, except that certain starting energies give
a better fit to the data.

The fact that they get more attraction
with the largest starting energy is rather simple. With a
positive $\omega$ we are closer to the poles in the
energy denominator of $G$, i.e.\
\begin{equation}
       \frac{1}{\omega - H_0}.
\end{equation}
However, the choice of a
positive starting energy is not straightforward
in the $G$-matrix calculation. With a negative starting energy
(appropriate for the low-lying states of finite nuclei),
there are no poles in the above energy  denominator. Actually,
a principle value integration should have been performed in the above
calculation of $G$. This is however not our main objection
against the use of the starting energy as a variable by ZBJVM.
With a multi-shell model space, one can no longer use a fixed starting
energy, rather, the starting energy should take into account
the fact
that the single-particle energies are no longer degenerate.
As an example, consider the matrix element
$\bra{(0d_{5/2})^2}G(\omega)\ket{(0d_{5/2})^2}$ coupled to
$JT=10$. This matrix element would enter our multi-shell
calculations in Table I. The correct starting energy should be
the unperturbed energy of two particles in the $d_{5/2}$ orbit.
This would correspond to $-44$ MeV in our example. This starting
energy gives a matrix element of $0.20$ MeV. In the previous section
we used a fixed starting energy of $-100$ MeV, which would give us
a matrix element of $0.48$ MeV. Thus, if the other matrix elements
behave in a similar way (and they do), the use of a
degenerate model space as done by ZBJVM, becomes meaningsless.
A scheme which takes
the starting energy dependence into account,
was recently proposed by Suzuki {\em et al.}
\cite{npa94}.

Thus, the starting energy is not a parameter which one can
choose in order to obtain a good correspondence with the data.
The dependence of $\omega$ must be taken into account
in the calculations. However, this leads us to our last point, namely
that of the non-hermiticity of the effective interaction.
We have in our calculations used a fixed starting energy, in order
to avoid this problem, which arises even at the level of the $G$-matrix.

\section{Non-hermiticity of the effective interaction}

In the first point we stressed the need of
including the ladder diagram of Fig.\ \ref{fig:fig2}.
However, if one does this, a more serious problem arises,
namely that of the non-hermiticity of the effective interaction.
Assume now that the intermediate states in the two-particle
ladder diagram are those of the $1s0d$-shell only.
Diagram (b) in Fig.\ \ref{fig:fig2} is then proportional
to
\begin{equation}
  -\frac{1}{4\hbar\Omega}\bra{(0p)^2}G\ket{(1s0d)}\bra{(1s0d)}G
   \ket{(0s)^2},
\end{equation}
where the intermediate states must be those of the $1s0d$-shell if
we use a model space for the effective interaction which consists
of the $0s$- and the $0p$-shells. $\Omega$ is the
oscillator frequency. The starting energy corresponds to the
unperturbed energy of $0s$-shell. If we now evaluate diagram (c),
we get
\begin{equation}
  -\frac{1}{2\hbar\Omega}\bra{(0s)^2}G\ket{(1s0d)}\bra{(1s0d)}G
   \ket{(0p)^2},
\end{equation}
which yields a strongly non-hermitian effective interaction. The starting
energy
corresponds here to the unperturbed energy of two particles in the
$0p$-shell. As done
by ZBJVM, one could ignore ladder diagrams in the definition of the
effective interaction, and thereby obtain a hermitian effective
interaction in terms of $G$ only\footnote{In our $G+2P$ calculation in Section
II
we used a fixed starting energy, in order to avoid the non-hermiticity. Even
with
the $G$-matrix, the effective interaction will be non-hermitian if we do not
use a fixed starting energy.}.

However, as discussed in Section II,
if one truncates the
model space, one has to include the ladder diagram, yielding
a non-hermitian interaction.
It is important to note that this non-hermiticity
arises only if we approximate RS perturbation theory to a given order.
If all terms are taken into account, this problem does not occur.
Viable approaches to obtain an order-by-order effective interaction
which is hermitian,
have recently
been proposed by Lindgren \cite{lind91} and Kuo {\em et al.}
\cite{kuo93}.

This strong non-hermiticity is also present if one includes
folded diagrams as well, as done in the recent work
of Jaqua {\em et al.} \cite{jhbv94}. The same critical remarks
in the above Sections apply to that work as well.

\section{Conclusion}

In summary, we have shown that
the differences between results for various
model spaces obtained by the authors of Ref.\ \cite{prc48},
is due to the omission of the particle-particle
ladder diagrams in their calculations. Thus, the conclusion
by ZBJVM, that the binding energies do not stabilize
as functions of various model spaces, is not correct.
Actually, we have demonstrated that low-order perturbation
theory gives the same results within a small model space as the
calculations in terms of the $G$-matrix in a large model space.
Moreover, we argue that the use of a fixed starting
energy by Zheng {\em et al.} may not be a
viable approach, since the
calculations involve several shells, and the starting
energy should take this into account.

However, if one wishes to properly  evaluate the starting
energy dependence and include the ladder diagrams,
one has to face the problem of the non-hermiticity
of the effective interaction, since one is dealing with
an interaction defined for several shells.

Finally, all such calculations, which involve many single-particle
orbits from several shells, do become prohibitevely time-consuming
for all nuclei but the lightest ones.
Thus, even with the present
increased computing power, the perspectives for computing properties
of
more interesting systems like $^{18}$O,
are rather meagre.

\clearpage

\begin{figure}
     \setlength{\unitlength}{1mm}
     \begin{picture}(50,50)
     \end{picture}
     \caption{Two different choices for the Pauli operator $Q$ used in the
     calculation of the $G$-matrix. See text for further discussion.}
     \label{fig:fig1}
\end{figure}

\begin{figure}
      \setlength{\unitlength}{1mm}
      \begin{picture}(50,50)
      \end{picture}
      \caption{(a) is the two-particle ladder diagram to second order
      in the $G$-matrix (wavy line).
      (b) and (c) are examples of ladder diagram contributions
      with intermediate states from the $1s0d$-shell only.}
      \label{fig:fig2}
\end{figure}

\begin{table}[hbtp]
\begin{center}
\caption{Eigenvalues for a system with two particles for
different model spaces. The second column lists the results
for a model space consisting of the single-particle orbits
of the $0s$-shell only. The third column includes
the $0p$-shell while the fourth and fifth columns include the
$1s0d$- and $1p0f$-shells, respectively. The row denoted by
$G$ means that only
the $G$-matrix defined in the text is used, while
$G+2P$ includes the two-particle ladder diagram to third
order in $G$. The results are scaled so that $\varepsilon_{0s_{1/2}}=0$.}
\begin{tabular}{rrrrr}
&&&&\\
&\multicolumn{1}{c}{$0s$}&
\multicolumn{1}{c}{$0s-0p$}&
\multicolumn{1}{c}{$0s-1s0d$}&
\multicolumn{1}{c}{$0s-1p0f$}\\
\hline
&&&&\\
\multicolumn{1}{c}{$JT^{\pi}=10^+$}&&&&\\
\multicolumn{1}{c}{$G$}&-7.83&-9.13&-11.03&-12.35\\
\multicolumn{1}{c}{$G+2P$}&-12.37&-12.33&-12.30&--\\
&&&\\
\end{tabular}
\end{center} \label{tab:tab1}
\end{table}


\begin{references}
\bibitem{prc48} D.\ C.\ Zheng, B.\ R.\ Barrett, L.\ Jaqua, J.\ P.\
Vary and R.\ J.\ McCarthy, Phys.\ Rev.\ C {\bf 48}, 1083 (1993)
\bibitem{kkko76}  E.M. Krenciglowa, C.L. Kung, T.T.S. Kuo and E. Osnes, Ann.
of Phys. {\bf 101}, 154 (1976)
\bibitem{mac89}  R. Machleidt, Adv. Nucl. Phys. {\bf 19}, 189 (1989)
\bibitem{npa94} K.\ Suzuki, R.\ Okamoto, T.\ T.\ S.\ Kuo
and P.\ J.\ Ellis, Nucl.\ Phys.\ {\bf A567}, 576 (1994)
\bibitem{lind91}I.\ Lindgren, J.\ Phys.\ B: At.\ Mol.\ Opt.\ Phys.\
{\bf 24}, 1143 (1991)
\bibitem{kuo93} T.\ T.\ S.\ Kuo, P.\ J.\ Ellis, J.\ Hao,
Z.\ Li, K.\ Suzuki and R.\ Okamoto, Nucl.\ Phys.\ {\bf A560}, 621 (1993)
\bibitem{jhbv94}L.\ Jaqua, B.\ R.\ Barrett, J.\ P.\
Vary and R.\ J.\ McCarthy, Nucl.\ Phys.\  {\bf A571}, 243 (1994)
\end{references}
\end{document}